\documentclass{iau} 
\usepackage{graphicx}

\title[Herschel Footprint Database and Service] %% give here short title %%
{Herschel Footprint Database and Service}

\author[Varga-Vereb\'{e}lyi$^1$ et al.]   %% give here short author list %%
{E. Varga-Vereb\'{e}lyi$^1$,
 L.~Dobos$^2$, T. ~Budav\'{a}ri$^3$  and Cs.~Kiss$^1$}

\affiliation{
$^1$ Konkoly Observatory of the Hungarian Academy of Sciences, Budapest, Hungary\\
$^2$ Department of Physics of Complex Systems, E\"{o}tv\"{o}s Lor\'{a}nd University, Budapest, Hungary\\
$^3$ Department of Applied Mathematics and Statistics, The Johns Hopkins University, Baltimore, Maryland, USA\\
}

\pubyear{2015}
\volume{xxx}  %% insert here IAU Symposium No.
\jname{Title of your IAU Symposium}
\editors{A.C. Editor, B.D. Editor \& C.E. Editor, eds.}
\begin{document}

\maketitle

\begin{abstract}
We created the Herschel\footnote{{\it Herschel} is an ESA space observatory with science instruments provided by European-led Principal Investigator consortia and with important participation from NASA.} Footprint Database and web services for the Herschel Space Observatory imaging data.
For this database we set up  a unified data model for the PACS and SPIRE Herschel instruments, 
from the pointing and header information of each observation, generated and stored sky coverages (footprints) of the observations
 in their exact geometric form. With this tool we extend the capabilities of the 
Herschel Science Archive by providing an effective search tool that is able to find observations 
for selected sky locations (objects), or even in larger areas in the sky. 


%% add here a maximum of 10 keywords, to be taken form the file <Keywords.txt>
\end{abstract}

\firstsection 
%\vspace*{-0.2cm}
\section{Introduction}
The Herschel Space Observatory (\cite{herschel}) was a cornerstone mission of the
European Space Agency, in operation between 2009 -- 2013. The two imaging 
instruments, the PACS \cite{Poglitsch2010} and SPIRE \cite{spire} photometers 
covered the far-infrared and submillimeter wavelengths (55 -- 672\,$\mu$m) and studied 
the formation of galaxies in the early Universe, their subsequent evolution and investigated the
formation of stars and their interaction with the interstellar medium.
All observations performed by Herschel are stored in the Herschel Science Archive (HSA) and accessible through the HSA User Interface\footnote{http://archives.esac.esa.int/hsa/ui/hui.jnlp} for the astronomical community.

We extended the capabilities of the HSA by providing a tool that can search for measurements (OBSIDs) based on sky coordinates, identify the area covered by specific measurements (footprint) and provide capabilities for effective astronomical queries. The Herschel Footprint Database 
was built using the solutions applied in previous footprint database services like 
the Sloan Digital Sky Survey (Budav\'ari et al., 2010). Our aim was to build a database and Herschel 
specific toolkit that is able to find each observation that included a selected sky 
position (or object) and its neighbourhood, or can identify observations that covered a certain 
area in the sky. 

To prepare our own footprint database, it was necessary to obtain the information 
on the exact bore-sight in any type of observation, irrespectively of the scientific 
content of the measurement. We created a unified 
data model for the PACS and SPIRE Herschel instruments from their originally different data structures
and downloaded the sky coverage information for each observation. 

\begin{figure}
%\vspace*{-0.3cm}
\includegraphics[width=0.328\textwidth]{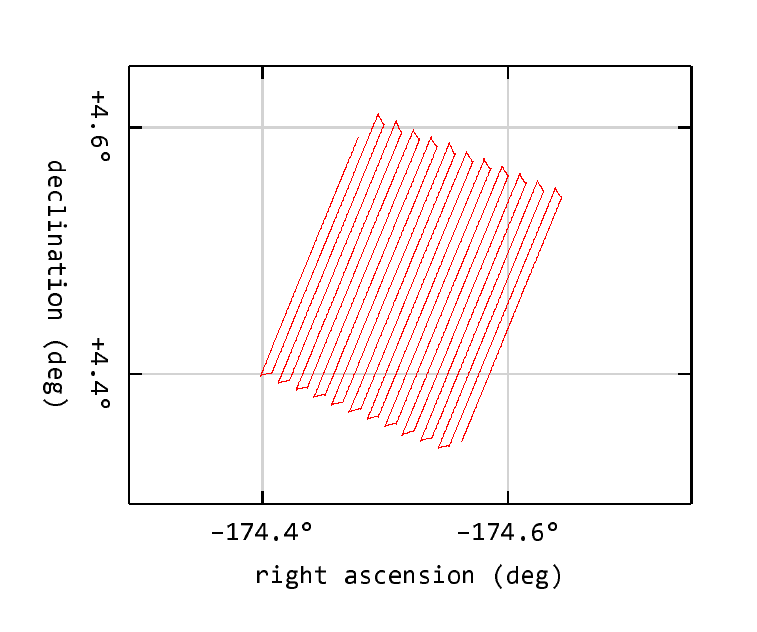}
\includegraphics[width=0.328\textwidth]{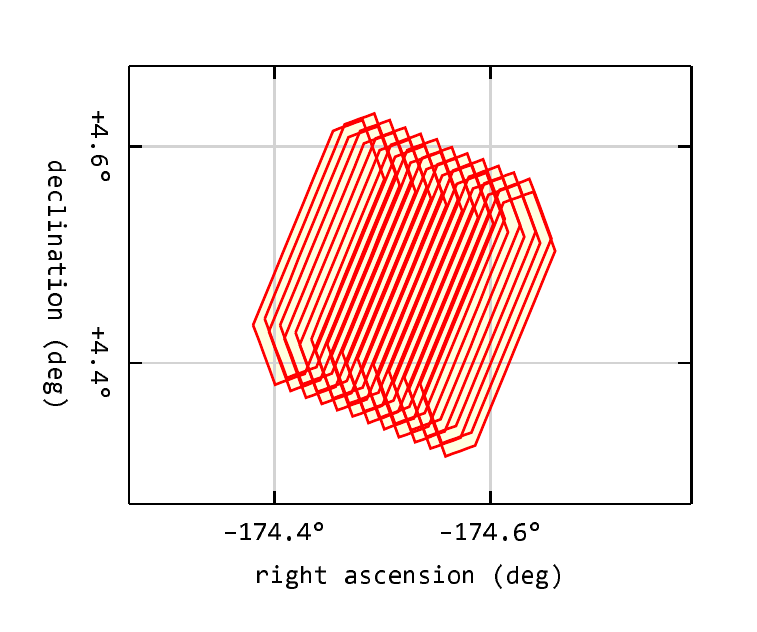}
\includegraphics[width=0.328\textwidth]{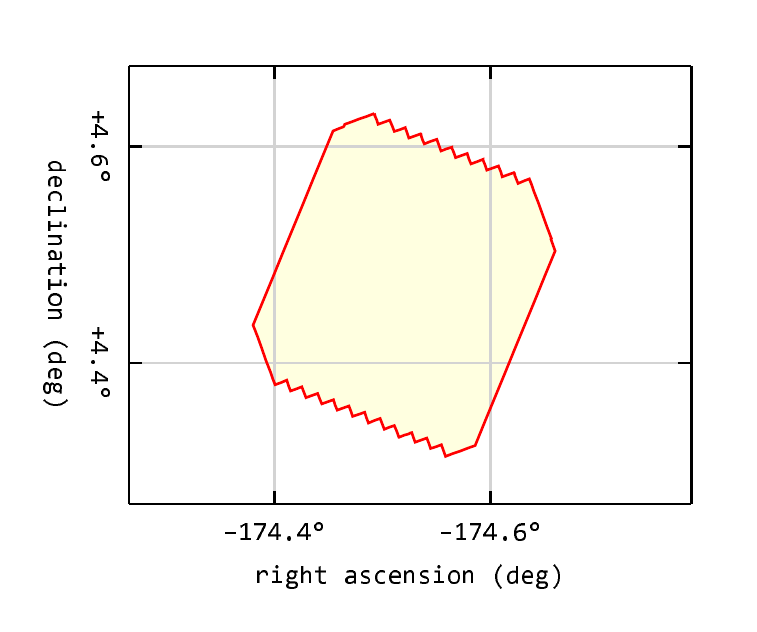}
\label{fig:pointing}
\caption{Reconstruction of the footprint from raw pointing data of PACS. Left panel: Raw pointing data, with telescope turn-arounds. These turn-arounds are filtered out and the entire scan curve is split into individual scan legs. Middle panel: Footprints of the individual, overlapping scan legs. Right panel: Final footprint, generated by taking the union of individual scan legs. Calculating the exact union by removing overlapping regions is necessary to determine the area of the footprint. }\hfill
\vspace*{-0.2cm}
\end{figure}

\vspace*{-0.5cm}
\section{Overview}

The unified data model was created using the pointing and header informations of all scan maps 
(PACS, SPIRE and PARALLEL mode observations) from the HSA.  
Scan maps in the HSA are usually processed at Level 2 or 2.5 which are the equivalent of flux-calibrated mosaic images in FITS format with a mask. Reconstructing the exact footprints of scan maps from these highly processed images would be challenging, mostly due to concave regions. Therefore, we had to reach back to initial pointing data - the pointing history of the telescope in a certain time window - and calculate footprints from the trajectories swept by the field of view of the various instruments.

%The sky coverage data are stored as non-pixelated, analytic description of the geometry. 
Sky coverage is stored in exact geometric form allowing for precise area calculations of scan legs and the entire map. 
To describe celestial  regions in an analytic form we upgraded the Spherical Library by 
\cite{Budavari2010} designed with astronomical use cases in mind. 
From the subsequent processing of the raw pointing information the generated footprint
database was entirely implemented in SQL. Figure~\ref{fig:pointing}. shows an example and explanation
of the reconstruction of a footprint from raw pointing data.

For a powerful and useful database the proper indexing is essential.
Orders of magnitude (or more) faster search can be done when we properly arrange indexing and 
linking our different kind of data. We used the HTM indices \cite{Kunszt2000} for the Herschel 
Footprint Database. We also created a visualization tool to display footprints with two type of methods 
to reduce the complexity of outlines.

With this database users can easily inspect if an area or a specific position in the sky was 
observed by the Herschel or not. The database calculate the exact area of the observations and visualise it. The database is accessible via a web site (http://herschel.vo.elte.hu) and also as a set of REST web service functions which makes it usable from program clients like Python or IDL scripts. Data will be available in various formats including Virtual Observatory standards.

\vspace*{-0.2cm}
\begin{acknowledgment}
The research was supported by the Hungarian  OTKA NN 114560 grants and PECS Contract Nr. 4000109997/13/NL/KML of the Hungarian Space Office and the ESA. 
\end{acknowledgment}

\end{document}